\documentclass[aps,pre,twocolumn,floatfix]{revtex4-1}

\usepackage{amsmath}
\usepackage{amssymb}
\usepackage{graphicx}
\usepackage{dcolumn}
\usepackage{bm}
\usepackage[usenames,dvipsnames]{xcolor}
\usepackage{multirow}
\usepackage{hyperref}
\usepackage{epstopdf}

\usepackage[utf8]{inputenc}
\usepackage[T1]{fontenc}
\usepackage{microtype}
\usepackage{txfonts}
\usepackage{ulem}
\usepackage{tabularx}

\usepackage{url,hyperref}
\hypersetup{colorlinks=true, linkcolor=blue, urlcolor=black, citecolor=blue}

\begin{document}


\title{Electrostatic pair-interaction of nearby metal or metal-coated colloids at fluid interfaces}

\author{Rick Bebon}
\author{Arghya Majee}
\email{majee@is.mpg.de}
\affiliation{Max Planck Institute for Intelligent Systems, Stuttgart, Germany}
\affiliation{IV. Institute for Theoretical Physics, University of Stuttgart, Germany}

\date{July 28, 2020}

\begin{abstract}
In this paper, we theoretically study the electrostatic interaction between 
a pair of identical colloids with constant surface potentials sitting in 
close vicinity of each other at a fluid interface. By employing a simplified 
yet reasonable model system, the problem is solved within the framework of 
classical density functional theory and linearized as well as nonlinear 
Poisson-Boltzmann (PB) theory. Apart from providing a sound theoretical 
framework generally applicable to any such problem, our novel findings, all 
of which contradict common beliefs, include the following: first, quantitative 
as well as qualitative differences between the interactions obtained within 
the linear and the nonlinear PB theories; second, the importance of the 
electrostatic interaction between the omnipresent three-phase contact 
lines in interfacial systems; and third, the occurrence of an attractive 
electrostatic interaction between a pair of identical metal colloids. The 
unusual attraction we report on largely stems from an attractive line 
interaction which although scales linearly with the size of the particle, 
can compete with the surface interactions and can be strong enough to 
alter the nature of the total electrostatic interaction. Our results 
should find applications in metal or metal-coated particle-stabilized 
emulsions where densely packed particle arrays are not only frequently 
observed but are sometimes required. 
\end{abstract}

\maketitle

\hypersetup{linkcolor=blue, urlcolor=blue, citecolor=blue}


\section{Introduction}\label{I}

Since its discovery in the early twentieth century by Ramsden \cite{Ram03} 
and Pickering \cite{Pic07}, stabilizing emulsions using colloidal particles 
instead of surfactants has become a standard technique. This new class of 
emulsions are popularly known as Pickering emulsions and finds application 
in diverse areas such as food \cite{Ray14, Ber15, Pic15}, cosmetic 
\cite{Ray14, Tan15, Mar16-1}, petroleum \cite{Uma18}, pharmaceutical or 
biomedical \cite{Son12, Lin13, Son13, He13}, and optical industries \cite{Ede16}, 
among others. Major advantages that make Pickering emulsions in general 
preferable over conventional surfactant-stabilized emulsions are its enhanced 
stability and reduced toxicity \cite{Wu16, Mar20}. Another important aspect 
is the availability of a wide range of particles differing in their composition 
and shape \cite{Bin17, Yan17, Gho19}. Among them, metal colloids are 
gaining increasing attention owing to their numerous applications facilitated 
by the advancement of opto-electronics and nanotechnology in the twenty-first 
century \cite{Sca18}.

Self-assembly of metal particles at fluid interfaces together with 
distinct optical, electrical, and catalytic properties of metal particles 
are exploited in liquid like mirrors \cite{Yoc03, Fan13}, sensors 
\cite{Sah12, Ede13, Pag14}, detectors \cite{Cec13}, filters \cite{Smi14}, 
antennas \cite{Ede13}, controllable and targeted drug delivery equipment 
\cite{Son12, Lin13, Son13, He13}, plasmonic rulers \cite{Tur12}, or in 
purification processes \cite{Cro10}. Due to their antimicrobial effect, 
silver particle stabilized emulsions are suitable for biomedical and 
textile applications \cite{Sam17}. For many of these applications, 
particularly for optical applications, a regular array of particles 
is needed which can be easily achieved at fluid interfaces. Contrary 
to the topological defects commonly observed for solid-liquid interfaces, 
the fluidic nature of these interfaces allows them to self-heal without 
any external influence, leading to strikingly uniform films spanning 
over large areas \cite{Cie10, Smi16}. Moreover, due to their defect-free 
nature such structures are easy to reproduce and being fluidic, they are 
easily deformable \cite{Smi16}.

Even a nanoparticle, while at fluid interfaces, typically feels a strong 
trapping potential several orders of magnitude larger than the thermal 
energy, originating from the reduction of the fluid-fluid interfacial area 
\cite{Bin00, Bin06, Boo15}. This restricts the motion of the particle to 
be only along the interfacial plane. Consequently, the arrangement of 
particles within a monolayer and the stability of the overall structure are 
largely dictated by their lateral interaction. For charged particles, one of 
the dominant contributions to the total lateral interaction usually comes 
from the electrostatic origin \cite{Oet08, Luo12, Kra16, Isa17}, which is 
the focus of the present study. 

If the inter-particle separations are large, then, from the electrostatic 
point of view, the system behaves like a set of interacting point dipoles 
as the counter-ion cloud surrounding each particle differs in the two 
adjacent fluid media and leads to an effective dipole normal to the 
interfacial plane \cite{Pie80}. The large separation distance also allows 
for a point-particle assumption and for an analytical solution using the 
linearized PB theory or the Debye-H\"uckel (DH) 
theory \cite{Hur85}. Whereas this simple yet useful picture has sparked 
the interest of a bunch of subsequent studies \cite{Kra16, Ave02, Nik02, 
Wue04, Fry07, Dom08, Mas10, Gao15, Pet16, Rah19}, little is known 
about the opposite limit, i.e., when the inter-particle separation is small 
compared to the size of the particles. Clearly, the dipolar assumption 
fails in this situation.

Most of the few recent efforts \cite{Maj14, Maj16, Lia16, Maj18-1, Maj18-2} 
targeted toward addressing the short inter-particle separations have 
considered particles with constant surface charge densities, which, at 
a simplistic level, describes dielectric or insulating particles the best 
\cite{Der16, Mar16-2}. Conductive metal particles, on the other hand, 
are characterized by constant surface potentials \cite{Mar16-2}. In a 
very recent publication \cite{Zig20}, the interaction between particles 
at an air-water interface with constant surface potentials only in the 
portions immersed in water is addressed using the linearized PB theory. 
However, equipotential metal particle surfaces carry the same constant 
potential irrespective of the adjacent fluid phase. Not only that, metal 
particle stabilized emulsions often feature significantly polarizable oils 
\cite{Yen14, Smi16, Sca18} which behave quite differently compared 
to air. Moreover, the use of the linearized PB theory at short separations 
may not be accurate as well \cite{Maj16}. Therefore, a proper description 
of the electrostatic interaction between a pair of metal particles situated 
very close to each other at a fluid interface is still lacking and we aim 
toward filling this gap here.

The short separation situation we consider is frequently encountered 
experimentally for metal particle-stabilized emulsions 
\cite{Ede16, Sca18, Pag14, Smi16, Rei04, Coh07, Yen14, Too16}. 
In fact, for nanoplasmonic mirrors or detectors, in order to get substantial 
signal, one needs densely packed array of relatively big particles 
($\gtrsim 40\,\mathrm{nm}$ in diameter) with average surface to 
surface distance being smaller than the particle radii \cite{Ede16, Sca18, 
Cec13, Fla10}. Not only that, our study could be equally useful for 
interfacial Janus colloids \cite{Fer15, Sto19, Die20} when the metal caps 
face each other or for core-shell particles consisting of metal shells with 
metallic or cost-effective non-metallic cores \cite{Pag14, Abi07}. One 
should also note that, strictly speaking, all particles including the metal 
ones in electrolyte solutions are charge-regulated \cite{Maj18-3, San19}. 
We do not account for such complexities here. However, the constant 
potential case should still remain insightful as the charge regulation solution 
is bounded by constant charge and constant potential limits \cite{Mar16-2}. 

\begin{figure}[!t]
\begin{center}
\includegraphics[width=1\columnwidth]{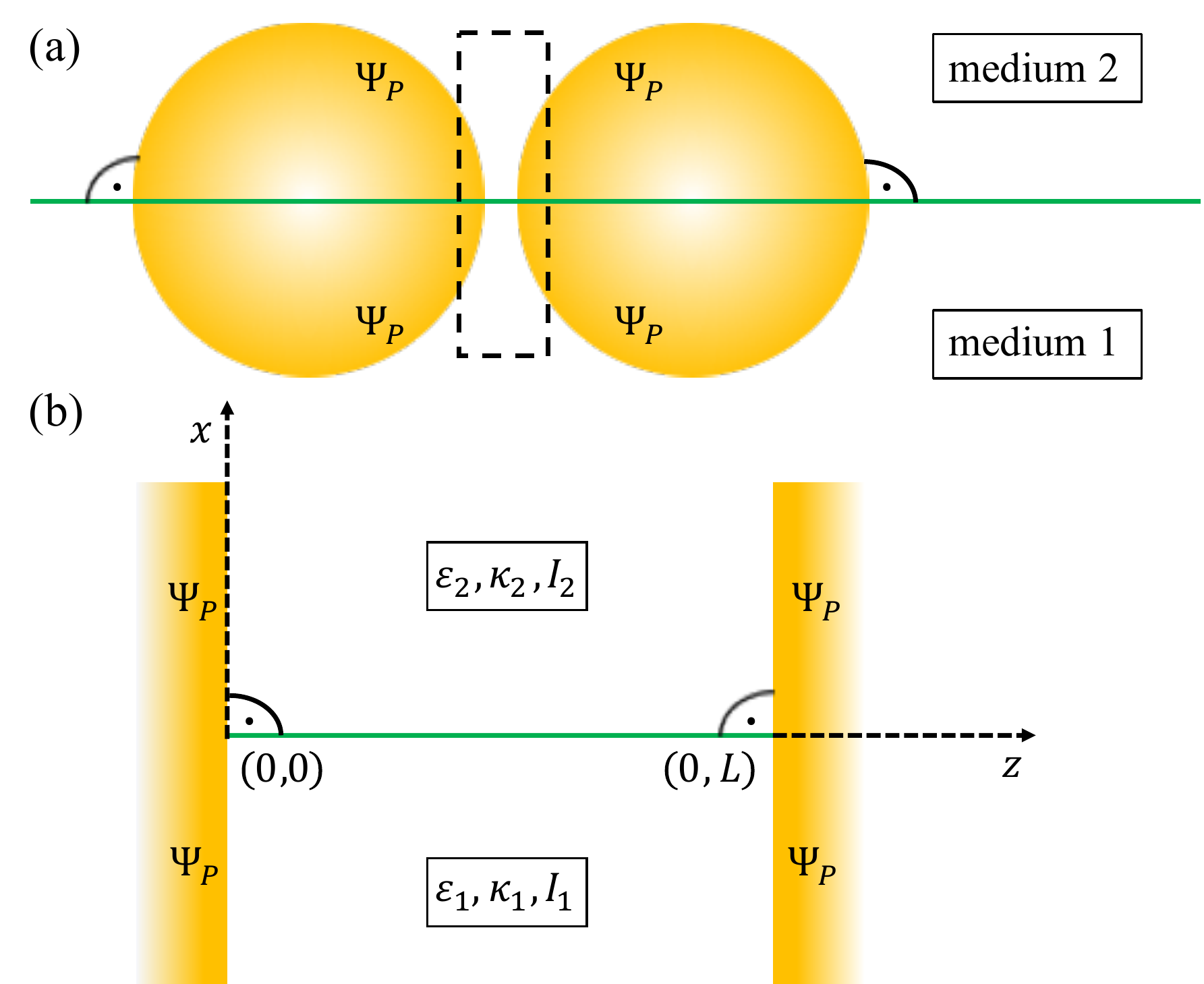}
\caption{(a) Graphical illustration of the system under consideration. 
         Two identical particles (represented by the yellow circles) 
         with constant surface potentials $\Psi_P$ are sitting next 
         to each other at a flat interface (indicated by the green 
         horizontal line) formed by two immiscible electrolytes 
         denoted as medium ``1''  (for $x<0$) and as medium ``2'' 
         (for $x>0$). The surface-to-surface separation between the 
         particles is small compared to their size and each particle 
         is submerged equally in the two fluid phases corresponding 
         to a liquid-particle contact angle of $90^\circ$. (b) Enlarged 
         view of a model system that approximates the boxed region in 
         part (a). Owing to their short separation, the particles are 
         assumed to be parallel flat plates situated at $z=0$ and 
         $z=L$ and aligned vertically with respect to the fluid 
         interface. Fluid phase  ``1'' (``2'') has dielectric constant
         $\varepsilon_1$ ($\varepsilon_2$). The ionic strength of the 
         added binary monovalent salt and the corresponding inverse 
         Debye length in medium ``1'' (``2'') are given by $I_1$ 
         ($I_2$), and $\kappa_1$ ($\kappa_2$), respectively.}
\label{Fig1}
\end{center}
\end{figure}

In order to tackle the problem efficiently, we use some justified approximations 
(see Fig.~\ref{Fig1}). First, the short inter-particle separation is exploited 
to treat the particles as flat plates by ignoring their curvatures. This assumption, 
motivated by the Derjaguin approximation, has been used before in this context 
as well \cite{Maj14, Maj16, Maj18-1}. Second, we assume the fluid-fluid 
interface to be flat as interfacial deformations are usually negligible for smooth 
particles up to a few micrometers in size \cite{Kra00, Sta00, Oet08, Ana16}. 
Moreover, we consider a $90^\circ$ liquid-particle contact angle which 
corresponds to the maximum reduction of interfacial area for any spherical 
particle and has been particularly shown to be pivotal for the entrapment of 
nanoparticles \cite{Dua04, Hu12}.  Several other existing theoretical models 
also rely on these latter two assumptions \cite{Dom08, Upp14, Nal14, Bos16, Zig20}.

The electrostatic problem for the resulting model system, as depicted 
in Fig.~\ref{Fig1}(b), is studied within the framework of classical density 
functional theory (DFT) which requires the free energy of the system as 
the only input. A subsequent minimization of this free energy then leads 
to the governing equation for the electrostatic potential along with the 
boundary conditions and the ensuing effective interaction, which 
automatically ensures self-consistency, can also be attained easily. We 
solve the problem separately using the linearized PB theory and the 
nonlinear PB theory. In each case, the effective interaction is decomposed 
into parts proportional to the surface areas of the plates and to the lengths 
of the three-phase contact lines. We also provide results for the 
separation-independent interactions, i.e., surface, interface, and line 
tensions present in the system. Apart from offering a comparison, the 
analytically obtained results within the linear theory also serve as 
checks for those obtained numerically within the nonlinear PB theory in 
proper limits. Our results feature several novel aspects that contradict 
common beliefs. First, we show that the interaction of the three-phase 
contact lines plays a crucial role in determining the nature of the total 
electrostatic interaction. This is not intuitively obvious as the line 
part to the total electrostatic interaction scales linearly with the size 
of the particle whereas the surface parts scale with the square of the 
size of the particle. Second, our findings suggest that, depending upon 
the specific system under consideration, the interaction obtained within 
the linearized PB theory can differ quantitatively as well as qualitatively 
from that obtained within the nonlinear PB theory. While the quantitative 
differences are not unexpected at small inter-particle separations, the 
qualitative one is indeed surprising. Third and as a most striking result, 
it is shown that the electrostatic interaction of \textit{identical} metal 
particles at fluid interface is not necessarily always repulsive. The 
unexpected attraction stems from the line interaction energy which can 
be repulsive as well as attractive and, depending upon the system, can 
be strong enough to alter the nature of the total electrostatic interaction. 

\section{Model and formalism}\label{II}

In a Cartesian coordinate system, the region of interest is the space 
$x\in(-\infty,\infty)$, $z\in[0,L]$ bounded by two plates positioned at $z=0$
and $z=L$ and filled with two immiscible fluid phases creating an interface
at $x=0$ (see Fig.~\ref{Fig1}(b)). As the plates mimic metal surfaces, they 
are equipotential. Moreover, they are considered to be identical and are 
modeled as constant potential surfaces, i.e., they carry the same surface 
potential $\Psi_P$ irrespective of the separation distance $L$ between 
them. The fluid phase occupying the space $x<0$ $(x>0)$ is denoted as 
medium ``1'' (``2''). The dissolved salt is a simple binary compound 
composed of oppositely charged monovalent ingredients only. Consequently, 
the ionic strength equals the salt concentration and its bulk value in medium 
$i\in\{1,2\}$ is given by $I_i$. The background solvents, i.e., the fluids are 
treated as structureless incompressible linear dielectrics with dielectric 
constants $\varepsilon_i = \varepsilon_{r,i}\varepsilon_0$, $i\in\{1,2\}$, 
where $\varepsilon_{r,i}$ is the relative permittivity of medium $i$ and 
$\varepsilon_0$ is the vacuum permittivity. Please note that the length 
scale of our interest is the Debye screening length 
$\kappa_i^{-1}=\sqrt{\varepsilon_{r,i}/\left(8\pi\ell_B I_i\right)}$, 
where $\ell_B=e^2/\left(4\pi\varepsilon_0 k_BT\right)$ denotes the vacuum 
Bjerrum length with the elementary positive charge $e$, the Boltzmann 
constant $k_B$, and the temperature $T$. Therefore, any phenomena 
occurring on a smaller length scale such as the scale of bulk correlation 
length or molecular length are discarded. This includes the structures 
formed in the liquids due to the presence of the salt ions or of the surfaces 
and the interface, and associated changes in the ion number density profiles 
$n_{\pm}\left(\mathbf{r}\right)$ \cite{Bie12}. Thus, both the ionic strength 
profile $I(\mathbf{r})$ and the dielectric constant profile $\varepsilon(\mathbf{r})$ 
vary steplike at the interface. However, we consider the variation in the local 
charge density $e\left[n_{+}\left(\mathbf{r}\right)-n_{-}\left(\mathbf{r}\right)\right]$ 
as it varies on the scale of the Debye length. On this length scale, the salt ions 
can be considered as point like objects and, following the standard practice 
within a mean-field theory, we ignore any ion-ion correlation. With all these, 
the grand canonical density functional in the units of the thermal energy 
$k_BT=1/\beta$ for our system, which is in equilibrium with the ion reservoirs 
provided by the bulk of the two fluid media, is given by
\begin{align}
   \beta\Omega\left[n_{\pm}\right] = 
    &\int\limits_Vd^3r\Bigg[\sum\limits_{k=\pm}n_k\left(\mathbf{r}\right)
     \left\{\ln\left(\frac{n_k\left(\mathbf{r}\right)}{\zeta_k}\right)-1
     +\beta V_k\left(\mathbf{r}\right)\right\}\nonumber\\
    &+\frac{\beta\mathbf{D}
     \left(\mathbf{r},\left[n_\pm\right]\right)^2}{2\varepsilon\left(\mathbf{r}\right)}\Bigg]\nonumber\\
    &+\beta\Psi_P\int\limits_{\partial V}d^2r
     \hat{\text{\boldmath$\nu$}}\left(\mathbf{r}\right)\cdot\mathbf{D}\left(\mathbf{r},\left[n_\pm\right]\right).
\label{eq:1}
\end{align}
In this expression, within the curly brackets, the first two terms correspond to 
the entropic ideal gas contribution of the ions with fugacities $\zeta_{\pm}$, 
the third term describes the contribution due to ion solvation expressed via 
an external potential $V_{\pm}\left(\mathbf{r}\right)$ acting on the ions. The 
term quadratic in the electric displacement vector $\mathbf{D}$ includes all 
the Coulomb electrostatic interactions in the system due to the presence of 
surface charges and ions, whereas the last term stands for the work done 
by the system to maintain the plates at a constant surfaces potential $\Psi_P$. 
The integration volume $V$ is the space $(x,z)\in(-\infty,\infty)\times[0,L]$ 
enclosed by the bounding surfaces $\partial V$, i.e., the two plates at $z=0$ 
and $L$ with $\hat{\text{\boldmath$\nu$}}\left(\mathbf{r}\right)$ denoting the 
unit outward normal to these surfaces. Although Eq.~\eqref{eq:1} is central 
to both the linearized and the nonlinear PB theory and what we do next is in 
principle the same within both the theories, the ways in which we proceed 
from here differ slightly as the former is analytically tackleable whereas one 
needs to resort to numerical techniques for the latter. Hence, below we 
discuss them separately.

\subsection{Linear theory}\label{II-A}

Within the linearized PB theory, one assumes that the deviations in the 
ion number densities compared to their bulk values are small. Consequently, 
$\Omega\left[n_{\pm}\right]$ is expanded in terms of these small deviations 
to obtain $\widetilde\Omega\left[n_{\pm}\right]$ by retaining terms up to 
quadratic order in the expansion. A minimization of 
$\widetilde\Omega\left[n_{\pm}\right]$ with respect to $n_{\pm}$ then 
provides the equilibrium profiles $n^{\text{eq}}_{\pm}$ which, together with 
the Gauss's law, leads to the linearized PB or DH equation
\begin{align}
   \Delta\Psi_i\left(\mathbf{r}\right)=\kappa_i^2\left[\Psi_i\left(\mathbf{r}\right)-\Psi_{b,i}\right]
\label{eq:2}
\end{align}
to be solved for the electrostatic potential $\Psi_i\left(\mathbf{r}\right)$ in 
medium $i\in\left\{1,2\right\}$ (see Chapter 2 of \cite{Beb18} for details). In 
conjunction with this, the following boundary conditions, which also come 
out of the minimization process, must be satisfied. First, both the electrostatic 
potential and the $x$-component of the electric displacement vector should be 
continuous at the fluid-fluid interface, i.e., at any given $z$-value, 
$\Psi\left(x=0^-,z\right)=\Psi\left(x=0^+,z\right)$, and 
$\varepsilon_1\partial_x\Psi_1\left(x=0^-,z\right)=\varepsilon_2\partial_x\Psi_2\left(x=0^+,z\right)$ 
should hold true. Second, the electrostatic potential must match the surface 
potential of the two plates at $z=0$ and $L$: 
$\Psi_i\left(x,z=0\right)=\Psi_i\left(x,z=L\right)=\Psi_P$, irrespective of the value 
of $x$. Moreover, the electrostatic potential needs to be finite while approaching 
$x\rightarrow\pm\infty$ which, in fact, is a prerequisite for using a Derjaguin-like 
approximation and is typically satisfied due to electrostatic screening.
$\Psi_{b,i}$ in Eq.~\eqref{eq:2} refers to the 
bulk electrostatic potential in medium $i\in\left\{1,2\right\}$. By construction, 
the bulk electrostatic potential profile $\Psi_b(\mathbf{r})$ reads as
\[
 \Psi_b(\mathbf{r}) = 
  \begin{cases} 
   \Psi_{b,1}=0&(\text{in medium ``1''}) \\
   \Psi_{b,2}=\Psi_D&(\text{in medium ``2''}).
  \end{cases}
\]
This difference ($\Psi_D$) arises from contrasting solvation energies of 
the ions in the two media and is known as the Donnan potential or Galvani 
potential difference \cite{Bag06}.

As shown in Sec.~\ref{III-A}, Eq.~\eqref{eq:2} is analytically solvable for 
our set-up. Once the electrostatic potentials $\Psi_i\left(\mathbf{r}\right)$ are 
known, they are used to calculate $n_{\pm}^{\text{eq}}[\Psi]$ by considering 
the ion density profiles as functionals of $\Psi$. Finally, inserting these 
$n_{\pm}^{\text{eq}}[\Psi]$ back into the expression for $\widetilde\Omega\left[n_{\pm}\right]$, 
one obtains the grand potential 
$\widetilde\Omega\left(L\right)=\widetilde\Omega\left[n_{\pm}\left[\Psi\right]\right]$ 
of our system which combines the following contributions distinctly different 
from each other according to their origin:
\begin{align}
   \widetilde\Omega\left(L\right)=&\sum\limits_{i\in\left\{1,2\right\}}\left[\Omega_{b,i}V_i
   +\left(\gamma_i+\omega_{\gamma,i}\left(L\right)\right)A_i\right]\nonumber\\
   &+\gamma_{1,2}A_{1,2}+\left(\tau+\omega_{\tau}\left(L\right)\right)\ell.
\label{eq:3}
\end{align}
Here, $\Omega_{b,i}=-2I_i/\beta$ is the osmotic or entropic energy 
contribution due to the ideal gas formed by the ions expressed per 
volume $V_i$ of medium $i\in\{1,2\}$, $\gamma_i$ is the surface tension 
acting between each plate and medium $i$, $\omega_{\gamma,i}(L)$ 
is the surfaces interaction energy density between the portions of the 
plates immersed in and acting through medium $i$, $A_i$ is the total 
surface area of the two plates immersed in medium $i$, $\gamma_{1,2}$ 
is the interfacial tension between medium ``$1$'' and medium ``$2$'', 
$A_{1,2}$ is the total area of the fluid-fluid interface; $\tau$ is the line 
tension present at the three-phase contact lines at $(0,0)$ and $(0,L)$, 
and $\omega_{\tau}(L)$ is the contribution due to interaction between 
the contact lines expressed per total length $\ell$ of the two contact lines. 
Clearly, $\gamma_i$ and $\tau$ are $L$-independent as they result 
from the interaction of a single plate with its adjacent fluid(s). For the 
purpose of calculating the effective interaction between two colloids 
as shown in Fig.~\ref{Fig1}(a), the terms involving $\Omega_{b,i}$ and 
$\gamma_{1,2}$ in Eq.~(\ref{eq:3}) also become irrelevant as the total 
volume of the fluids and the total area of the fluid-fluid interface, which 
not only include the boxed region in Fig.~\ref{Fig1}(a) but also the outer 
space, do not change while changing the separation distance $L$ 
between the colloids. Therefore, the effective inter-particle interaction, 
which is what we are interested in, is exclusively dictated by the contributions 
involving the surface interaction energy density $\omega_{\gamma,i}(L)$ 
and the line interaction energy density $\omega_{\tau}(L)$. Please note that 
they are constructed such that $\omega_{\gamma,i}(L\rightarrow\infty)\rightarrow0$ 
and $\omega_{\tau}(L\rightarrow\infty)\rightarrow0$.

\subsection{Nonlinear theory}\label{II-B}

Within the nonlinear PB theory, the density functional in Eq.~\eqref{eq:1} is 
directly minimized with respect to $n_{\pm}$ which provides the equilibrium 
profiles $n_{\pm}^{\text{eq}}$ for the ion number densities. Using these profiles, 
one finally ends up getting the nonlinear PB equation
\begin{align}
   \Delta\left(\beta e\Psi_i\left(\mathbf{r}\right)\right)=\kappa_i^2\sinh\left[\beta e\left(\Psi_i\left(\mathbf{r}\right)-\Psi_{b,i}\right)\right]
\label{eq:4}
\end{align}
along with the same boundary conditions as listed below Eq.~\eqref{eq:2} to 
be satisfied by the electrostatic potential $\Psi_i\left(\mathbf{r}\right)$ in medium 
$i\in\left\{1,2\right\}$. In principle, after solving Eq.~\eqref{eq:4}, the grand potential 
$\Omega\left(L\right)$ can also be obtained from Eq.~\eqref{eq:1} by inserting the 
equilibrium ions profiles $n_{\pm}^{\text{eq}}\left[\Psi\right]$. However, in order to 
facilitate the whole process, we use the density functional 
\begin{align}
   \beta\Omega_{\text{fem}}\left[\Psi\right] = 
  &\int\limits_{V}d^3r\Bigg[2I(\mathbf{r})\cosh\left\{\beta e \left(\Psi(\mathbf{r})-\Psi_b(\mathbf{r})\right)\right\}\nonumber\\
  &+\frac{\beta\varepsilon(\mathbf{r})}{2}\left\{\left(\partial_x\Psi(\mathbf{r})\right)^2+\left(\partial_z\Psi(\mathbf{r})
    \right)^2\right\}\Bigg]
\label{eq:5}
\end{align}
instead and obtain the electrostatic potential $\Psi\left(\mathbf{r}\right)$ 
(defined as $\Psi\left(\mathbf{r}\right)=\Psi_1\left(\mathbf{r}\right)$ for 
$x<0$ and $\Psi\left(\mathbf{r}\right)=\Psi_2\left(\mathbf{r}\right)$ for 
$x>0$) by minimizing it numerically using Rayleigh-Ritz-like \textit{f}inite 
\textit{e}lement \textit{m}ethod \cite{Bra10}. Please note that the minimization of this 
functional with respect to $\Psi\left(\mathbf{r}\right)$ also results in the 
PB equation [Eq.~\eqref{eq:4}] together with all the boundary conditions that 
need to be satisfied. Once the potential $\Psi\left(\mathbf{r}\right)$, which 
minimizes Eq.~\eqref{eq:5} is obtained, it can be inserted back in 
Eq.~\eqref{eq:5} to obtain $\Omega_{\text{fem}}^{\text{min}}\left(L\right)$ 
which, at equilibrium, is related to the grand potential $\Omega\left(L\right)$ 
by $\Omega\left(L\right)=-\Omega_{\text{fem}}^{\text{min}}\left(L\right)$. 
Without loss of generality, we first rewrite Eq.~\eqref{eq:5} by expanding the 
function $\cosh\left\{\beta e \left(\Psi(\mathbf{r})-\Psi_b(\mathbf{r})\right)\right\}$ 
in a Taylor series,
\begin{align}
   \beta\Omega_{\text{fem}}^{(m)}\left[\Psi\right] = 
  &\int\limits_{V}d^3r\Bigg[2I(\mathbf{r})\sum\limits_{k=0}^{m}\frac{\left\{\beta e \left(\Psi(\mathbf{r})-\Psi_b(\mathbf{r})\right)\right\}^{2k}}{\left(2k\right)!}\nonumber\\
  &+\frac{\beta\varepsilon(\mathbf{r})}{2}\left\{\left(\partial_x\Psi(\mathbf{r})\right)^2+\left(\partial_z\Psi(\mathbf{r})
    \right)^2\right\}\Bigg],
\label{eq:6}
\end{align}
where $m$ can be interpreted as the degree of nonlinearity. For $m=1$, one recovers 
the linearized PB problem whereas $m\rightarrow\infty$ corresponds to the full nonlinear 
problem. We increase $m$ in Eq.~\eqref{eq:6} step by step starting from $m=1$. For 
each $m$-value, $\Omega^{(m)}\left(L\right)=-\Omega_{\text{fem}}^{(m)\text{min}}\left(L\right)$ 
is calculated and further decomposed into different interaction parameters as described 
in Eq.~\eqref{eq:3}, 
\begin{align}
   \Omega^{(m)}\left(L\right)=-\Omega_{\text{fem}}^{(m)\text{min}}\left(L\right) = 
   &\sum\limits_{i\in\left\{1,2\right\}}\left[\Omega_{b,i}V_i+\left(\gamma_i+\omega_{\gamma,i}\left(L\right)\right)A_i\right]\nonumber\\
   &+\gamma_{1,2}A_{1,2}+\left(\tau+\omega_{\tau}\left(L\right)\right)\ell,
\label{eq:7}
\end{align}
with all the quantities in the right hand side carrying the same meaning 
as defined for Eq.~\eqref{eq:3}. Numerical extraction of these contributions 
requires solving some subproblems with the following set-ups: (i) single 
fluid medium ``1'' (``2'') spanning the lower (upper) half-space in 
Fig.~\ref{Fig1}(b) in the absence of any plates; the corresponding grand 
potential density $\Omega_{b,i}=-2I_i/\beta$ is easily obtained by simply 
setting $\Psi(\mathbf{r})=\Psi_b(\mathbf{r})$ and $\Psi_P=0$ in 
Eq.~(\ref{eq:6}), (ii) only the two fluid media present in the absence of 
any plates, (iii) a single plate present in contact with a single semi-infinite 
fluid medium, (iv) two plates interacting across a single fluid medium, and 
(v) a single plate touching two semi-infinite fluids in its one side. A sequential 
subtraction of the grand potentials obtained for these subproblems then 
enables one to separate all the contributions in Eq.~(\ref{eq:7}). For $m=1$, 
the resulting interaction parameters are checked against those obtained 
analytically within the linear theory. With increasing $m$, for all the different 
system parameters considered here, we observe that 
$\Omega^{(m)}\left(L\right)=-\Omega_{\text{fem}}^{(m)\text{min}}\left(L\right)$ 
saturates for $m\geq4$. Therefore, in what follows, we call the results for 
$m=4$ as the solutions under the nonlinear theory. This is in accordance 
with the conclusion of Ref.~\cite{Maj16} and is not unexpected since we 
deal with the potentials of similar order of magnitude in both studies.

\section{Results and discussion}\label{III}

\subsection{Linear theory}\label{III-A}

\subsubsection{Electrostatic potential}\label{III-A-1} 

Our goal is to find the electrostatic potential which satisfies the DH 
equation (Eq.~\eqref{eq:2}) and the associated boundary conditions 
as mentioned below Eq.~\eqref{eq:2} for the system depicted in 
Fig.~\ref{Fig1}(b). To achieve this, we first dissect the original problem 
into the following subproblems: (i) two plates with constant surface 
potentials $\Psi_P$ interacting across fluid ``1'' or fluid ``2'' with bulk 
potential $\Psi_{b,1}$ or $\Psi_{b,2}$, respectively; the resulting 
potential in both cases depend on $z$-coordinate only, (ii) two fluids 
phases occupying the two half spaces $(x\gtrless0)$ forming an interface 
at $x=0$ in the absence of any plates; the corresponding electrostatic 
potential, which is solely $x$-dependent, and the electric displacement 
vector are continuous at the interface. Exploiting the linear nature of 
the governing equation, i.e., the DH equation, if we simply add the 
solution for medium $i$ in subproblem (i) with the corresponding ones 
obtained in subproblem (ii), the sums are also solutions of the DH 
equation and they satisfy most of the boundary conditions except for 
the continuity of the total electrostatic potentials at the interface. In 
order to get rid of this inconsistency we then seek for a correction 
function which also satisfies the DH equation and which, when added 
to the previously obtained sums, meets the remaining continuity condition 
without affecting any of the conditions that are already fulfilled. Such a 
function can easily be constructed by means of Fourier series expansion 
(for details, please see Chapter 3 of Ref.~\cite{Beb18}). With all these, 
the final expressions for the electrostatic potentials in medium $i\in\{1,2\}$ 
read as
\begin{align}
 \Psi_i(x,z\,;L)=&\Psi_{b,i}+\left(\Psi_P-\Psi_{b,i}\right)\frac{\sinh\left(\kappa_iz\right)-
              \sinh\left(\kappa_i\left(z-L\right)\right)}{\sinh\left(\kappa_iL\right)}\nonumber\\
             &-\sum\limits_{j\in\left\{1,2\right\}}^{j\neq i}\sum\limits_{n=1,3,\dots}^{\infty}\left(-1\right)^i
              4n\pi\varepsilon_jp_{n,j}(L)\xi_n(L)\nonumber\\
             &\times e^{-p_{n,i}(L)|x|}\sin\left(\frac{n\pi z}{L}\right),
\label{eq:8}
\end{align}
where 
\begin{align*}
 \xi_n(L)=\frac{\frac{\Psi_D}{n^2\pi^2}-\frac{\Psi_P}{n^2\pi^2+\kappa_1^2L^2}
          +\frac{\left(\Psi_P-\Psi_D\right)}{n^2\pi^2+\kappa_2^2L^2}}
          {\varepsilon_1p_{n,1}(L)+\varepsilon_2p_{n,2}(L)}
\end{align*}
and $p_{n,i}(L)=\sqrt{\frac{n^2\pi^2}{L^2}+\kappa_i^2}$. The first two 
terms in Eq.~\eqref{eq:8} together is the solution of the subproblem (i), i.e., 
potential distribution for two plates with constant potentials $\Psi_P$ interacting 
across medium $i$. The remainder represents the correction function. The 
solution of the subproblem (ii) is canceled by the first term ($n=0$) of the 
series representing the correction function. From Eq.~\eqref{eq:8}, one can 
trivially check that $\Psi_i(z=0)=\Psi_i(z=L)=\Psi_P$. In the limit $x\rightarrow\pm\infty$, 
the series term vanishes because of the exponential function and one is left 
with the first two terms describing the potential due to two plates with 
constant surface potential $\Psi_P$ interacting across medium $i$ with bulk 
potential $\Psi_{b,i}$. This is exactly the situation far away from the interface. 
Upon approaching simultaneously the limits $x\rightarrow\pm\infty$ and 
$L\rightarrow\infty$, the series term vanishes and the second term reduces to 
an exponentially varying function characteristic of a single plate placed in 
contact with an electrolyte solution. In addition, if we set $z\rightarrow\infty$ in 
Eq.~\eqref{eq:8}, we recover the bulk potential $\Psi_{b,i}$. As a side remark, 
we note that in the limit $L\rightarrow\infty$ only, the second term reduces to 
an exponentially decaying potential and the sum can be converted to an integral 
over $q=n\pi/L$. The resulting expression corresponds to the potential due to a 
single plate with surface potential $\Psi_P$ placed in contact with two immiscible 
fluids spanning the regions $x\gtrless0$ with $z>0$. This can also be derived 
independently with the help of Fourier transforms (for details, please refer to 
the derivation of Eqs.~(3.115) and (3.116) in Ref.~\cite{Beb18}).
\medskip\smallskip

\subsubsection{Interaction energies}\label{III-A-2}

With the electrostatic potential at hand, we can use it to calculate the grand potential 
$\widetilde\Omega\left(L\right)=\widetilde\Omega\left[n_{\pm}\left[\Psi\right]\right]$ 
and to derive the expressions for the interaction parameters of our system. As 
defined in Eq.~\eqref{eq:3}, this is achieved by distinguishing the terms proportional 
to $V_i$, $A_i$, $A_{1,2}$, and $\ell$. Moreover, the $L$-independent interactions 
are recognized in the terms proportional to $A_i$ and $\ell$ by taking the limit 
$L\rightarrow\infty$ (for details, please see Chapter 4 of Ref.~\cite{Beb18}).

The surface tensions, as defined in Eq.~\eqref{eq:3}, acting between each surface 
and its adjacent fluid medium $i\in\{1,2\}$ is given by
\begin{align}
  \gamma_i=-\frac{\varepsilon_i\kappa_i}{2}\left(\Psi_P-\Psi_{b,i}\right)^2,
\label{eq:9}
\end{align}
and the surface interaction energy between two surfaces with total area $A_i$ in 
medium $i\in\{1,2\}$ is given by
\begin{align}
 \omega_{\gamma,i}(L)=\frac{\varepsilon_i\kappa_i\left(\Psi_P-\Psi_{b,i}\right)^2}{2}
 \left[1+\frac{1-\cosh\left(\kappa_iL\right)}{\sinh\left(\kappa_iL\right)}\right].
\label{eq:10}
\end{align}
Whereas the amplitudes of both $\gamma_i$ and $\omega_{\gamma,i}(L)$ 
depend on $\varepsilon_i$, $\kappa_i$, $\Psi_P$, and $\Psi_{b,i}$, the decay 
rate of $\omega_{\gamma,i}(L)$ is uniquely governed by $\kappa_i$. In the 
limit of vanishing separation between the plates, Eq.~\eqref{eq:10} predicts 
a finite, repulsive interaction: 
$\omega_{\gamma,i}(L\rightarrow0)=\frac{\varepsilon_i\kappa_i}{2}\left(\Psi_P-\Psi_{b,i}\right)^2$. 
Contrary to a constant charge boundary condition, this non-divergent behavior 
in the $L\rightarrow0$ limit is characteristic of a constant potential boundary 
condition \cite{Mar16-2}. In the large asymptotic limit, Eq.~\eqref{eq:10} provides
$\omega_{\gamma,i}(L\rightarrow\infty)\simeq\varepsilon_i\kappa_i\left(\Psi_P-\Psi_{b,i}\right)^2\left[e^{-\kappa_iL}-\frac{e^{-2\kappa_iL}}{2}\right]$. 
Therefore, as dictated by the leading order term, the surface interaction 
$\omega_{\gamma,i}(L)$ decays monotonically $\sim e^{-\kappa_iL}$ for 
large separations between the plates. 

The interfacial tension per total interfacial area $A_{1,2}$ acting between the 
two fluid media reads as
\begin{align}
 \gamma_{1,2}=-\frac{\varepsilon_1\varepsilon_2\kappa_1\kappa_2\Psi_D^2}{2\left(\varepsilon_1\kappa_1
              +\varepsilon_2\kappa_2\right)}.
\label{eq:11}
\end{align}
As expected, the interfacial tension depends only on the properties of the two 
fluid media and not on any plate properties. Consequently, the expression for 
$\gamma_{1,2}$ does not differ from what one obtains for plates with constant 
surface charge densities \cite{Maj18-1}.

The line tension acting at both the three-phase contact lines expressed per 
total length of the two contact lines is given by
\begin{widetext}
\begin{align}
 \tau=&-\frac{\varepsilon_1\varepsilon_2\left(\Psi_P-\Psi_D\right)}{\pi}\int\limits_{0}^{\infty}dq
       \frac{p_1(q)}{p_2(q)}q^2\xi(q)
      +\frac{\varepsilon_1\varepsilon_2\Psi_P}{\pi}\int\limits_{0}^{\infty}dq\frac{p_2(q)}{p_1(q)}q^2\xi(q)\nonumber\\
      &-\frac{\varepsilon_1\varepsilon_2\Psi_D}{\pi}\int\limits_0^{\infty}
       \frac{dq}{\varepsilon_1p_1(q)+\varepsilon_2p_2(q)}
       \left[\frac{\left(\kappa_2^2-\kappa_1^2\right)\Psi_D}{p_1(q)p_2(q)+\frac{\kappa_1}{\kappa_2}p_2(q)^2}\right.
      \left.-\frac{\left(\kappa_2^2-\kappa_1^2\right)\Psi_P}{p_1(q)p_2(q)}
       -\frac{\kappa_1\kappa_2\Psi_D}{\varepsilon_1\kappa_1+\varepsilon_2\kappa_2}
       \left(\frac{\varepsilon_1}{p_1(q)+\kappa_1}+\frac{\varepsilon_2}{p_2(q)+\kappa_2}\right)\right],
\label{eq:12}
\end{align}
\end{widetext}
with
\begin{align*}
 \xi(q)=\frac{\frac{\Psi_D}{q^2}-\frac{\Psi_P}{q^2+\kappa_1^2}+\frac{\left(\Psi_P-\Psi_D\right)}{q^2+\kappa_2^2}}
        {\varepsilon_1p_1(q)+\varepsilon_2p_2(q)}
\end{align*}
and $p_{i}(q)=\sqrt{q^2+\kappa_i^2}$, and the line interaction energy (defined in Eq.~\eqref{eq:3}) acting 
between the two three-phase contact lines is given by
\begin{align}
 \omega_{\tau}(L)=&-2\varepsilon_1\varepsilon_2\left(\Psi_P-\Psi_D\right)\sum\limits_{n=1,3,\dots}^{\infty}
                   \xi_n(L)\frac{n^2\pi^2}{L}\frac{p_{n,1}(L)}{p_{n,2}(L)}\nonumber\\
                  &+2\varepsilon_1\varepsilon_2\Psi_P\sum\limits_{n=1,3,\dots}^{\infty}\xi_n(L)\frac{n^2\pi^2}{L}
                   \frac{p_{n,2}(L)}{p_{n,1}(L)}\nonumber\\
                  &-2\varepsilon_1\varepsilon_2\Psi_D\sum\limits_{n=1,3,\dots}^{\infty}\xi_n(L)p_{n,1}(L)p_{n,2}(L)L\nonumber\\
                  &+\frac{\varepsilon_1\varepsilon_2\kappa_1\kappa_2\Psi_D^2L}{4\left(\varepsilon_1\kappa_1+\varepsilon_2\kappa_2\right)}-\tau,
\label{eq:13} 
\end{align}
with $\xi_n(L)$ and $p_{n,i}(L)$ as defined below Eq.~\eqref{eq:8} and 
the expression for $\tau$ given in Eq.~\eqref{eq:12}. In the limit of 
vanishing separation ($L\rightarrow0$) between the plates, all the terms 
in Eq.~\eqref{eq:13} vanish except for the last term, i.e., $\tau$. Therefore, 
in this limit $\omega_{\tau}(L)$ stays finite unlike what one observes 
for constant charge boundary condition \cite{Maj14, Maj18-1}. Using the 
relation $\sum\limits_{n=1,3,\dots}^{\infty}\frac{1}{n^2}=\frac{\pi}{8}$ it 
is trivial to see that the terms $\propto L$, i.e., the third and the fourth 
terms in Eq.~\eqref{eq:13} cancel each other in the large asymptotic limit 
($L\rightarrow\infty$). The rest of the terms together lead to an overall 
exponential decay of the line interaction $\omega_{\tau}(L)$ in this limit. 
A striking observation regarding the line interaction for constant surface 
charge density boundary condition is that it is independent of the Donnan 
potential $\Psi_D$ \cite{Maj14, Maj18-1}. On the contrary, as Eq.~\eqref{eq:13} 
suggests, $\omega_{\tau}(L)$ does depend on $\Psi_D$ while using a 
constant surface potential boundary condition. This is indeed reasonable 
since the line part essentially captures the effects of the interface to the 
total inter-plate interaction and the contrast in the bulk potential should 
get reflected in it.

\subsection{Nonlinear theory}\label{III-B}

In this subsection, we discuss the results within the nonlinear 
theory and compare them with those obtained using the linear theory 
in Sec.~\ref{III-A}. In order to present our numerically obtained results 
and to gain insight into the variations of different parameters efficiently, 
we use dimensionless quantities. Accordingly, $\omega_{\gamma,i}(L)$ and 
$\omega_{\tau}(L)$ are expressed as $\beta\omega_{\gamma,i}(L)/\kappa_1^2$ 
and $\beta\omega_{\tau}(L)/\kappa_1$, respectively and their behavior is 
studied as functions of the scaled separation $\kappa_1L$. In doing so, the 
number of dimensionless free parameters reduces to only four: $\beta e\Psi_P$, 
$\beta e\Psi_D$, $I=I_2/I_1$, and $\varepsilon=\varepsilon_2/\varepsilon_1$. 
As the permittivities of oils used in typical experimental systems widely vary 
\cite{Yen14, Smi16, Sca18}, we consider two experimental systems differing 
significantly in their fluid contents for our discussion. The variations of different 
system parameters with respect to one of these systems are also presented 
which allow one to infer what could be expected to happen for an arbitrary 
general system.

\begin{figure}[!t]
\begin{center}
\includegraphics[width=7.0cm]{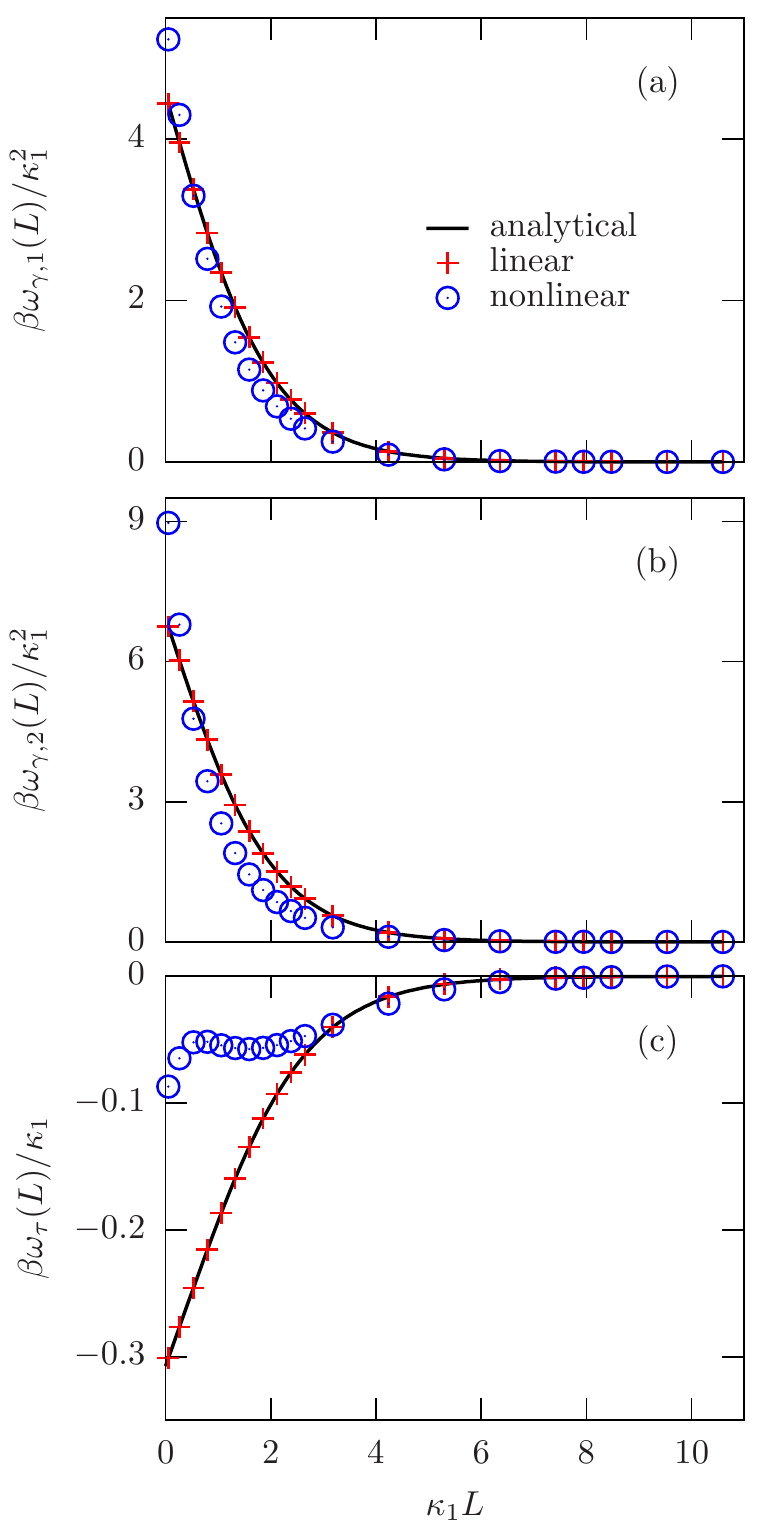}
\caption{(a) - (b) Variations of the surface interaction energy densities 
              $\omega_{\gamma,1}(L)$ and $\omega_{\gamma,2}(L)$, both expressed 
              in units of $\kappa_1^2/\beta$ and (c) variation of the line interaction 
              energy density $\omega_{\tau}(L)$ expressed in units of $\kappa_1/\beta$
              as functions of the scaled separation $\kappa_1L$ between the two surfaces 
              for standard set of parameters ($\beta e\Psi_P=-3$, $\beta e\Psi_D=1$, 
              $\varepsilon=62/72$, and $I=0.85$) corresponding to particles trapped at a 
              water--lutidine interface. As shown by all the plots, the numerically obtained 
              results for $m=1$ in Eq.~\eqref{eq:7} match perfectly with the analytically 
              obtained results within the linearized PB theory. Whereas both $\omega_{\gamma,1}(L)$ 
              and $\omega_{\gamma,2}(L)$ are repulsive and decay monotonically with 
              increasing separation distance $\kappa_1L$ under linearized as well as 
              nonlinear PB theory, the linear theory underestimates them at very 
              short separations and overestimates at larger separations. On the other 
              hand, the linear theory fails to qualitatively predict the correct variation 
              for $\omega_{\tau}(L)$. Whereas $\omega_{\tau}(L)$ is attractive everywhere 
              in the linear case, it can be attractive as well as repulsive within the nonlinear 
              theory.}
\label{Fig2}
\end{center}
\end{figure}

\subsubsection{Water--lutidine system}\label{III-B-1}

As our first example, we take a system with particles at a water--lutidine 
(2,6-dimethylpyridine) interface at temperature $T=313\,\mathrm{K}$ 
above the critical temperature \cite{Gra93}. The water-rich phase is 
denoted as the medium ``1'' and the lutidine-rich phase as the medium 
``2''. The dissolved salt is NaI with bulk ionic concentrations 
$I_1=1\,\mathrm{mM}$ and $I_2=0.85\,\mathrm{mM}$ which leads to 
a Donnan potential estimated to be $1\,k_BT/e$ \cite{Ine94, Bie12}. The 
relative permittivities of the two fluids are $\varepsilon_{r,1}=72$ and 
$\varepsilon_{r,2}=62$ \cite{Ram57, Lid02}. Typically metal surfaces 
are negatively charged in an electrolyte solution \cite{Smi17}. Therefore, 
the particles are considered to be carrying a constant potential 
$\Psi_P=-3\,k_BT/e\approx-77\,\mathrm{mV}$. Please note that the 
surface potential is expected to be slightly higher than the zeta potentials 
usually measured in experiments. Therefore, the dimensionless parameters 
of our system, which we call as the standard system, are $I=0.85$, 
$\varepsilon=62/72$, $\beta e\Psi_D=1$, and $\beta e\Psi_P=-3$.

\begin{figure}[!t]
\begin{center}
\includegraphics[width=7.0cm]{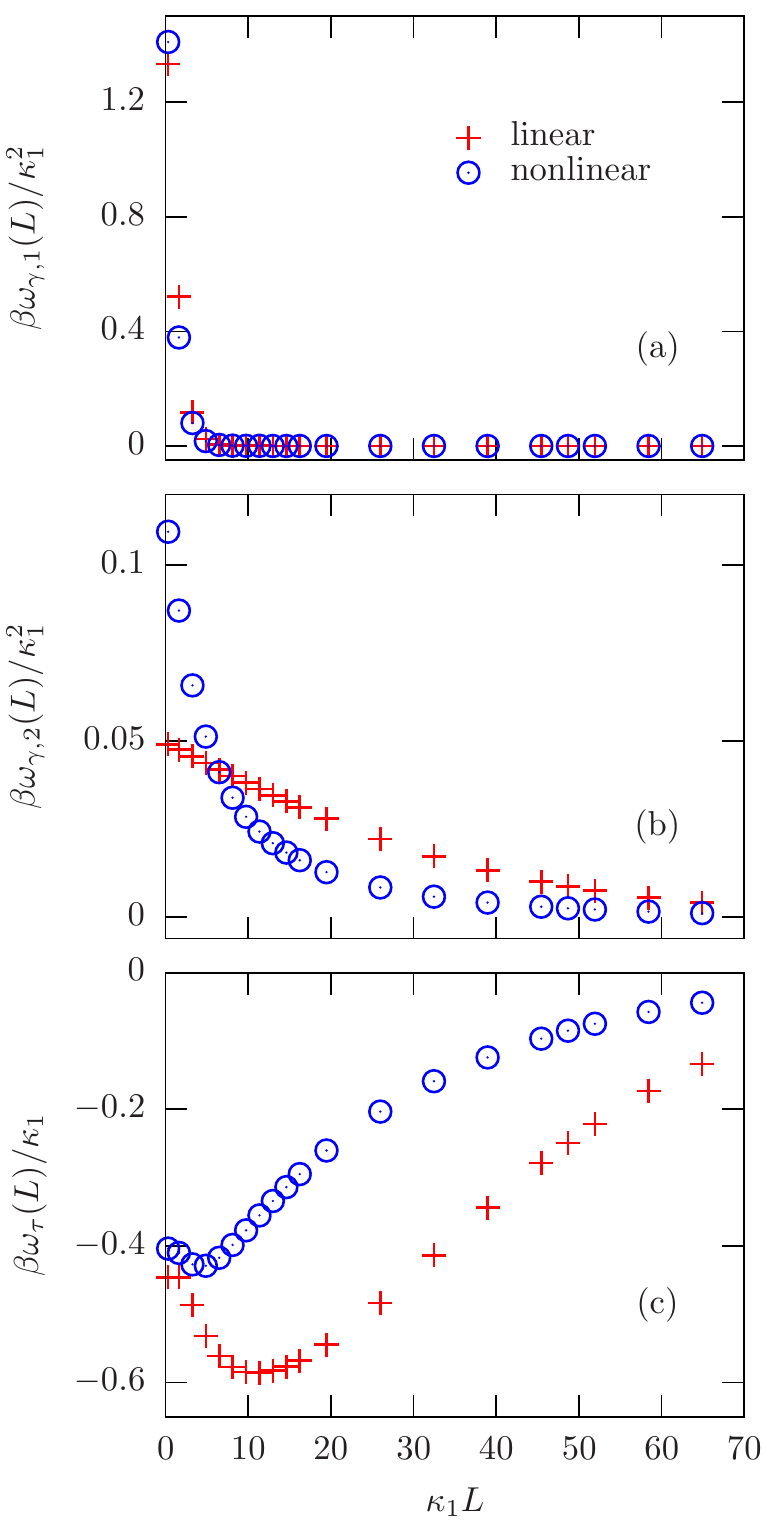}
\caption{(a) - (b) Variations of the surface interaction energy densities 
              $\omega_{\gamma,1}(L)$ and $\omega_{\gamma,2}(L)$, both expressed 
              in units of $\kappa_1^2/\beta$ and (c) variation of the line interaction 
              energy density $\omega_{\tau}(L)$ expressed in units of $\kappa_1/\beta$
              as functions of the scaled separation $\kappa_1L$ between the two surfaces. 
              The parameters ($\beta e\Psi_P=-3$, $\beta e\Psi_D=3.8$, $\varepsilon=10.3/80$, 
              and $I=2.9\times10^{-4}$) used for the plots correspond to a system consisting 
              of particles trapped at a water--octanol interface. As shown by all the plots, 
              whereas both $\omega_{\gamma,1}(L)$ and $\omega_{\gamma,2}(L)$ are 
              repulsive and decay monotonically with increasing separation distance 
              $\kappa_1L$ under linearized as well as nonlinear PB theory, the linear 
              theory underestimates them at very short separations and overestimates 
              at larger separations. Regarding the variation of the line interaction 
              $\omega_{\tau}(L)$, whereas the linear theory qualitatively predicts the correct 
              behavior, it is inaccurate in predicting the magnitude of $\omega_{\tau}(L)$ 
              and the location of the minimum.}
\label{Fig3}
\end{center}
\end{figure}

The ensuing interaction energies between the two plates are presented 
in Fig.~\ref{Fig2}. For all the plots, the data points correspond to the 
numerically obtained results and the solid lines represent the analytically 
obtained expressions in Subsection \ref{III-A-2}. Figure~\ref{Fig2}(a) 
shows the variation of the scaled surface interaction energy density 
$\omega_{\gamma,1}(L)$ acting between the portions of the plates in 
contact with medium ``1'' as function of the varying scaled separation 
$L$. As one can see, both the linear and the nonlinear theory predict 
a monotonically decaying repulsive interaction between the plates, and 
as expected, the numerically obtained results corresponding to $m=1$ 
in Eq.~\eqref{eq:7} match perfectly with the analytically obtained solution 
within the linear theory throughout the range of separations considered 
here. While the results within the two theories qualitatively agree with 
each other, the linear theory underestimates the interaction at very short 
separations and overestimates at relatively larger separations. This 
discrepancy gradually drops down as the separation distance is increased. 
Similar conclusions can be drawn from Fig.~\ref{Fig2}(b) concerning the 
variation of the scaled surface interaction energy density $\omega_{\gamma,2}(L)$ 
acting between the portions of the plates in contact with medium ``2'' as 
function of the varying scaled separation $L$ within the two theories. The 
magnitudes of $\omega_{\gamma,1}(L)$ and $\omega_{\gamma,2}(L)$ 
are of the same order as the two fluids do not significantly differ in their 
properties and the surface potentials of the plates are also the same 
in both media. However, the surface interaction in medium ``2'' is slightly
stronger compared to medium ``1'' due to less screening. Figure~\ref{Fig2}(c) 
shows the variation of the scaled line interaction energy $\omega_{\tau}(L)$ 
as function of the scaled separation $L$ within both the theories. As one 
can infer, within the range of separations considered, the linear theory 
predicts a monotonically decaying attractive interaction with increasing 
separation distance between the plates. However, within the nonlinear 
theory the line interaction shows a qualitatively different behavior. It 
changes non-monotonically upon increasing the separation distance, 
shows a maximum at $\kappa_1L\approx1$ followed by a minimum at 
around $\kappa_1L\approx2$ and then decays monotonically. Clearly, 
the interaction is repulsive in between the maximum and the minimum. 
Please note that both the maximum and the minimum occur at distances 
much larger than the molecular length scale (typically a few angstroms).
Not only that, up to around $\kappa_1L\approx2$, the linear theory 
significantly overestimates the strength of the line interaction as well.

\subsubsection{Water--octanol system}\label{III-B-2}

Water--lutidine falls in a class of systems where the bulk properties, i.e., 
the relative permittivities and the bulk ionic strengths, and consequently 
the Debye screening lengths vary little in the two fluid phases. Although 
such combination of immiscible fluids are used in experiments 
\cite{Smi17, Sca18}, there is another frequently used category of systems 
where moderately polarizable oils are used \cite{Sca18}. Therefore, as 
a second example, we consider a system consisting of particles trapped 
at a water--octanol interface. At room temperature $T=300\,\mathrm{K}$ 
these two fluids are characterized by significantly different relative 
permittivities with $\varepsilon_{r,1}=80$ (for water as medium ``1'') and 
$\varepsilon_{r,2}=10.3$ (for octanol as medium ``2''). The ion-partitioning 
at this interface results in highly contrasting bulk ionic strengths as well. 
For $I_1=10\,\mathrm{mM}$ in the water phase, $I_2=2.9\times10^{-3}\,\mathrm{mM}$
in the octanol phase, leading to a Donnan potential $\Psi_D\approx3.8\,k_BT/e$. 
The associated inverse Debye lengths in the two phases are given by 
$\kappa_1\approx0.324\,\mathrm{nm^{-1}}$ and $\kappa_2\approx0.015\,\mathrm{nm^{-1}}$. 
The particles are considered to be carrying the same surface potentials 
as before, i.e., $\Psi_P=-3\,k_BT/e\approx-77\,\mathrm{mV}$. Hence, 
the dimensionless parameters of this system are given by $I=2.9\times10^{-4}$, 
$\varepsilon=10.3/80$, $\beta e\Psi_D=3.8$, and $\beta e\Psi_P=-3$.

\begin{figure}[!t]
\begin{center}
\includegraphics[width=8.0cm]{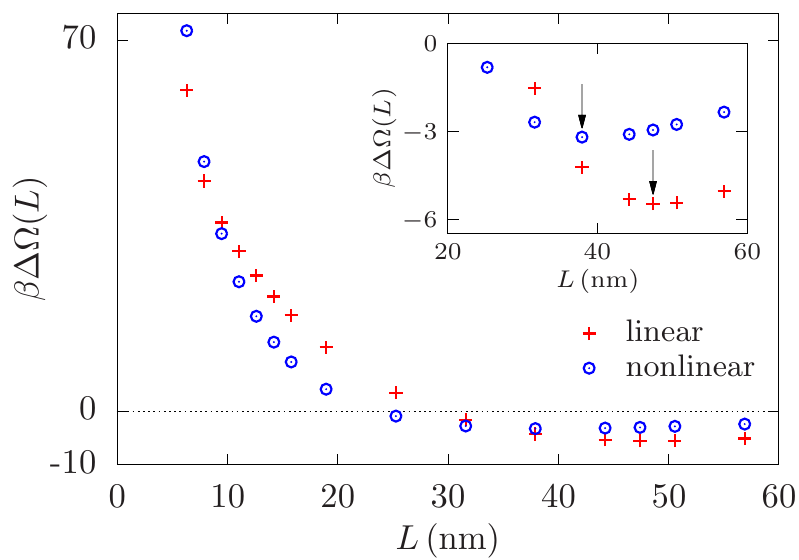}
\caption{Variation of total interaction energy 
         $\Delta\Omega(L)=\omega_{\gamma,1}(L)A_1+\omega_{\gamma,2}(L)A_2+\omega_{\tau}(L)\ell$ 
         between the two surfaces expressed in the units 
         of $1/\beta$ as a function of the separation $L$ 
         for a water--octanol system charecterised by 
         $\beta e\Psi_P=-3$, $\beta e\Psi_D=3.8$, $\varepsilon=10.3/80$, 
         $I_2=2.9\times10^{-2}\,\mathrm{mM}$, $I_1=0.1\,\mathrm{M}$, 
         $A_2=A_1=2573\,\mathrm{nm}^2$, and $\ell=114\,\mathrm{nm}$. 
         The total effective surface areas $A_i$ and the total 
         effective length of the three-phase contact lines 
         are rough estimates for $60\,\mathrm{nm}$ particles.
         As one can see, the interaction energy decreases 
         initially, shows a minimum, and finally increases 
         monotonically within both the linear (red crosses) 
         and the nonlinear PB theories (blue open circles). 
         The initial decreases in the interaction energy 
         corresponds to a positive effective force at 
         short separations $L$, implying that the interaction 
         is repulsive there. Beyond the position of the 
         minimum, it turns into an attractive one. Whereas 
         this feature remains the same within the two 
         theories, the location of the minimum (as indicated 
         by the arrows in the inset) shifts from 
         $L\approx 48\,\mathrm{nm}$ to a significantly 
         shorter separation $L\approx 38\,\mathrm{nm}$ 
         while calculated using the nonlinear PB theory. 
         Therefore, the particles will come much closer 
         than what is predicted by the linear theory.}
\label{Fig4}
\end{center}
\end{figure}

The resulting interactions for this system are shown in Fig.~\ref{Fig3}. 
Both $\omega_{\gamma,1}(L)$ and $\omega_{\gamma,2}(L)$, as shown 
in Figs.~\ref{Fig3}(a) and \ref{Fig3}(b), respectively predict 
monotonically decaying repulsive interactions under the linear as well 
as the nonlinear theory. But, as before, the linear theory underestimates 
the interactions at very short separations whereas overestimates them 
at relatively larger separations. Depending upon the fluid medium and the 
separation distance $L$, this mismatch can be significant (see, for example, 
the data presented in Fig.~\ref{Fig3}(b)). Except for very short separations, 
the interaction in medium ``2'', i.e., $\omega_{\gamma,2}(L)$ is stronger 
compared to that in medium ``1'' (presented by $\omega_{\gamma,1}(L)$). 
This is due to a weaker screening in the medium ``2'' compared with 
medium ``1'' which is evident from the rapid decay of $\omega_{\gamma,1}(L)$ 
in Fig.~\ref{Fig3}(a). At very short separations, however, one enters into 
the electric double layer and with decreasing separations, the ions, which 
are strongly attracted to the surfaces, need to be removed. As the ionic 
strength is higher in medium ``1'', the system needs to remove more ions 
leading to a stronger repulsion compared to medium ``2''. Concerning the 
line interaction energy $\omega_{\tau}(L)$, as shown in Fig.~\ref{Fig3}(c), 
it decays non-monotonically with increasing separation $L$ between the 
two plates. Initially its magnitude increases implying a repulsive interaction, 
which, after the occurrence of a minimum, turns into an attractive interaction. 
Although the linear theory predicts the qualitative behavior correctly, it 
significantly overestimates the strength of the interaction over the entire 
range of separations considered here and also fails to accurately locate 
the position of the minimum. It is worth mentioning that the line interaction 
for water--octanol system is much stronger compared to that for the 
water--lutidine system owing to greater contrast between the combination 
of the fluids used. While comparing the magnitudes in Figs.~\ref{Fig2} and 
\ref{Fig3}, please note that $\kappa_1\approx0.324\,\mathrm{nm^{-1}}$ for 
the water--octanol system whereas $\kappa_1\approx0.106\,\mathrm{nm^{-1}}$ 
for the water--lutidine system.

\begin{figure*}[!t]
\begin{center}
\includegraphics[width=14cm]{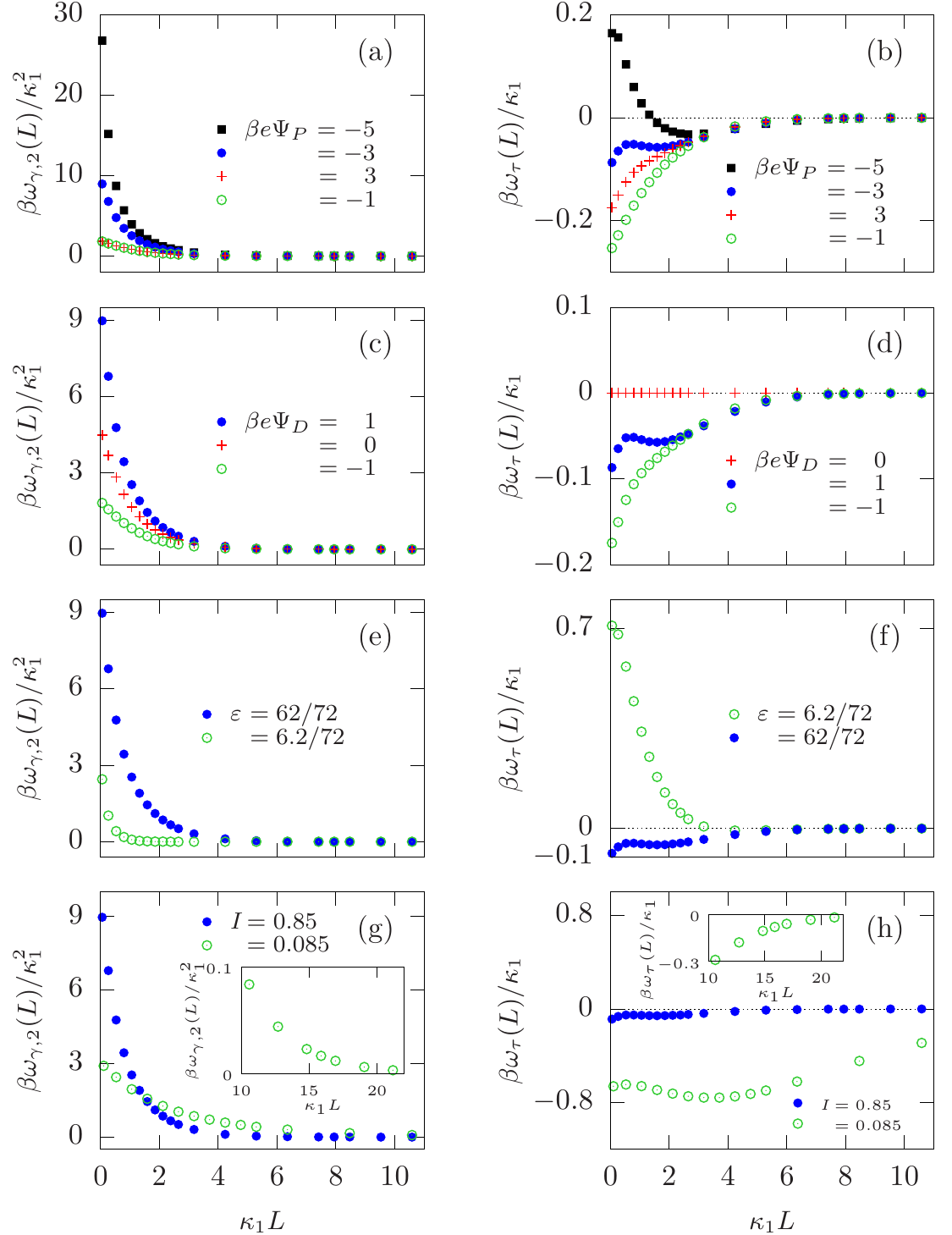}
\caption{Variations of the surface interaction energy density 
         $\omega_{\gamma,2}(L)$ expressed in units of $\kappa_1^2/\beta$ 
         (left panels) and variations of the line interaction energy 
         density $\omega_{\tau}(L)$ expressed in units of $\kappa_1/\beta$ 
         (right panels) as functions of the scaled separation $\kappa_1L$ 
         between the two plates for various sets of free dimensionless 
         parameters $\beta e\Psi_P$ (panels (a) and (b)), $\beta e\Psi_D$ 
         (panels (c) and (d)), $\varepsilon$ (panels (e) and (f)), and $I$ 
         (panels (g) and (h)). In all cases, only a single parameter 
         has been changed at a time from the standard water--lutidine 
         system and the interactions are compared. The parameters that 
         are changed are mentioned in each plot. As shown by all the 
         plots in the left panel, $\omega_{\gamma,2}(L)$ is always 
         repulsive and it increases with increasing difference 
         $\beta e\left|\Psi_P-\Psi_D\right|$ or $\beta e\Psi_D$ 
         or $\varepsilon$. On the other hand, whereas an 
         increase in $I$ strengthens $\omega_{\gamma,2}(L)$ 
         at very short separations, it weakens $\omega_{\gamma,2}(L)$ 
         at relatively larger separations. As shown by panel (b), 
         $\omega_{\tau}(L)$ shows a maximum followed by a 
         minimum for strongly or moderately negative surface 
         potentials. On the other hand, for positively charged 
         or weakly negatively charged particles $\omega_{\tau}(L)$ 
         remains attractive everywhere. With increasing values 
         of $\beta e\Psi_D$, $\omega_{\tau}(L)$ changes from 
         overall attractive to a non-monotonically varying 
         behavior showing a maximum at very short separations 
         followed by a minimum at larger separations. For 
         $\beta e\Psi_D=0$, $\omega_{\tau}(L)$ almost vanishes 
         as the contrast between the two fluid media reduces 
         in this case. With decreasing $\varepsilon$, the height 
         of the maximum of $\omega_{\tau}(L)$ increases and it 
         shifts to the left whereas the minimum becomes shallow 
         and shifts to larger distances. On the other hand, 
         with decreasing $I$, whereas the position of the 
         maximum does not change, the minimum shifts to larger 
         distances and the magnitude of $\omega_{\tau}(L)$ 
         increases. Insets of panels (g) and (h) show the 
         variations at larger separation distances as the 
         Debye screening length is larger for the smaller 
         $I$ values.}
\label{Fig5}
\end{center}
\end{figure*}

So far we have presented results concerning all the 
different parts, i.e., $\omega_{\gamma,1}(L)$, 
$\omega_{\gamma,2}(L)$ and $\omega_{\tau}(L)$ to the 
total electrostatic interaction. This indeed allows 
one to investigate the system in a detailed way and 
gain better insight from a theoretical point of view. 
We are also not required to restrict ourself to fixed 
particle sizes. However, experimentally it might be 
challenging to disentangle all these individual 
interaction parameters. Also, one might reasonably 
wonder about the relative importance of the line 
interaction energy $\omega_{\tau}(L)$ in comparison 
with the surface interaction energies $\omega_{\gamma,1}(L)$ 
and $\omega_{\gamma,2}(L)$. As Eq.~\eqref{eq:3} or 
\eqref{eq:7} suggests, the surface contributions 
to the total grand potential are proportional to 
the surface areas of the plates whereas the line 
contribution scales with the length of the three-phase 
contact lines. Therefore, for large surfaces, the 
surface contributions dominate anyway in due course. 
But what happens for typical system parmeters? To 
address these concerns, we present in Fig.~\ref{Fig4} 
the total interaction energy 
$\Delta\Omega(L)=\omega_{\gamma,1}(L)A_1+\omega_{\gamma,2}(L)A_2+\omega_{\tau}(L)\ell$ 
(i.e., the total $L$-dependent part in Eq.~\eqref{eq:3} 
or \eqref{eq:7}) for a water--octanol system with 
$\beta e\Psi_P=-3$, $\beta e\Psi_D=3.8$, 
$\varepsilon=10.3/80$, $I_2=2.9\times10^{-2}\,\mathrm{mM}$, 
$I_1=0.1\,\mathrm{M}$, $A_2=A_1=2573\,\mathrm{nm}^2$, 
and $\ell=114\,\mathrm{nm}$. The effective areas 
$A_i$ and and effective length of the three-phase 
contact lines are rough estimates for $60\,\mathrm{nm}$ 
particles. The quantity, i.e., $\Delta\Omega(L)$ 
we plot in Fig.~\ref{Fig4} is related to the effective 
electrostatic force $F(L)=-\partial\left(\Delta\Omega\right)/\partial L$ 
which can be considered as experimentally more relevant 
quantity. As the plot suggests, the total interaction 
energy $\Delta\Omega(L)$ initially decreases with 
increasing separation, reaches a minimum (marked 
with the arrows in the inset of Fig~\ref{Fig4}) and 
then increases within both the linear and the 
nonlinear PB theories. This implies a repulsive 
interaction till the position of the minimum and 
an attractive interaction beyond that. As both the 
surface parts are repulsive in Fig.~\ref{Fig3} 
throughout the range of separations $\kappa_1L$ 
considered here, the attractions clearly come from 
the line parts. Although both the linear and the 
nonlinear theory predict an attractive interaction 
beyond certain separations, the equilibrium separation 
(given by the position of the minimum) decreases 
significantly from $L\approx 48\,\mathrm{nm}$ 
to $L\approx 38\,\mathrm{nm}$ while calculated using 
the nonlinear PB thoery. It is worth mentioning that 
the line interaction starts to be comparable in 
magnitude with the surface parts at much shorter 
distances. Therefore, depending on the system, 
the line contribution can indeed dominate over 
the surface contributions even for typical system 
parameters. We also note that, as discussed in  
Subsection \ref{III-B-3} and as can be seen in 
Figs.~\ref{Fig5}(a) and \ref{Fig5}(b), the minimum 
can be moved to smaller separations and its depth 
can be increased by reducing the surface potential 
$\beta e\Psi_P$ as the magnitudes of the surface 
parts reduce whereas that of the line part increases 
with decreasing surface potential. 

\subsubsection{Variation of system parameters}\label{III-B-3}

Now we discuss the effects of changing the free parameters, i.e., 
$\beta e\Psi_P$, $\beta e\Psi_D$, $\varepsilon$, and $I$, of our system 
within the nonlinear PB theory. The comparisons are done with respect to 
the standard (water--lutidine) system. Please note that, the 
permittivity ratio $\varepsilon=\varepsilon_2/\varepsilon_1$ and the ionic 
concentration ratio $I=I_2/I_1$ can vary due to changes in the respective 
quantities in either or both fluids. However, the resulting total interaction 
is sensitive only to the ratios $\varepsilon$ and $I$. For our analysis, we 
have chosen to vary $\varepsilon$ and $I$ by changing $\varepsilon_2$ 
and $I_2$, respectively while keeping $\varepsilon_1$ and $I_1$ fixed. Therefore, 
only the surface interaction $\omega_{\gamma,2}(L)$ in medium ``2'' and 
$\omega_{\tau}(L)$ changes due to such variations but $\omega_{\gamma,1}(L)$ 
remains unaltered. $\omega_{\gamma,1}(L)$ is also independent of the Donnan 
potential $\beta e\Psi_D$. However, it changes due to variation in $\beta e\Psi_P$ but 
this happens in a fashion similar to what one observes for $\omega_{\gamma,2}(L)$. 
Therefore, here we show only the variations of $\omega_{\gamma,2}(L)$ and 
$\omega_{\tau}(L)$ upon changing the system parameters. 

As one can see from Fig.~\ref{Fig5}(a), the surface interaction $\omega_{\gamma,2}(L)$ 
in medium ``2'' increases with absolute value of the increasing contrast between 
surface potential $\Psi_P$ and the bulk potential $\Psi_{b,2}$. Consequently, no 
change is observed while varying $\beta e\Psi_P$ from $3$ to $-1$ (please note 
that $\beta e\Psi_{b,2}=1$ here). Similarly, $\omega_{\gamma,2}(L)$ increases 
with increasing Donnan potential $\beta e\Psi_D$ and the permittivity ratio 
$\varepsilon$; see Figs.~\ref{Fig5}(c) and \ref{Fig5}(e). Concerning the variation 
of $\omega_{\gamma,2}(L)$ upon increasing the ionic strength ratio $I$, 
$\omega_{\gamma,2}(L)$ increases at very short separations whereas it 
decreases at relatively larger separations; see Fig.~\ref{Fig5}(g). At these larger 
separations, with increasing ionic concentration, electrostatic screening increases. 
Therefore, the effective electrostatic interaction diminishes. However, at very short 
separations, within the electric double layer, the number of ions strongly attracted 
to the surfaces is more for higher ionic strengths. Consequently, more work needs 
to be done in order to decrease the separation distance as it requires removing 
these ions. This results in higher values for $\omega_{\gamma,2}(L)$. Please 
note that irrespective of the specific values of the system parameters 
$\omega_{\gamma,2}(L)$ is always repulsive.

The variations of $\omega_{\tau}(L)$ with different system parameters are 
shown in the right panel of Fig.~\ref{Fig5}. As one can see from Fig.~\ref{Fig5}(b), 
$\omega_{\tau}(L)$ shows a maximum followed by a minimum for strongly or 
moderately negative surface potentials. Therefore, the line interaction 
can be repulsive as well as attractive depending upon the separation distance 
$L$. On the other hand, for positively charged or weakly negatively charged 
particles the line interaction $\omega_{\tau}(L)$ remains attractive everywhere 
and decays monotonically. With increasing values of $\beta e\Psi_D$, 
$\omega_{\tau}(L)$ changes from overall attractive monotonically decreasing 
behavior to a non-monotonically varying behavior showing a maximum at 
very short separations followed by a minimum at larger separations; see 
Fig.~\ref{Fig5}(d). A special situation is presented by $\beta e\Psi_D=0$ which 
leads to an almost vanishing $\omega_{\tau}(L)$ as the contrast between 
the two fluid media reduces in this case. With decreasing $\varepsilon$, 
the height of the maximum of $\omega_{\tau}(L)$ increases and it shifts to 
the left whereas the minimum becomes shallow and shifts to larger distances; 
see Fig.~\ref{Fig5}(f). On the other hand, with decreasing $I$, whereas the 
position of the maximum does not change, the minimum shifts to larger 
distances and the magnitude of $\omega_{\tau}(L)$ increases; see Fig.~\ref{Fig5}(h).

\begin{table*}
\caption{Values of the surface tensions $\gamma_1$ (third column) and $\gamma_2$ 
              (fourth column) resulting from interactions of the plates across medium ``1'' and 
              medium ``2'', respectively, the interfacial tension $\gamma_{1,2}$ (fifth column) 
              acting between the two fluids, and the line tension $\tau$ (last column) acting 
              at the three-phase contact lines for all the systems considered in this study (listed 
              in the first column). The first block represents the results for our standard 
              water--lutidine system and the last block provides results corresponding to 
              the water--octanol system. The intermediate blocks provide results for 
              system where a single parameter has been changed at a time from the 
              standard system. These systems are denoted by the parameter that is changed. 
              The surface and the interfacial tension values are given in the units of 
              $k_BT/\mathrm{nm}^2$ while the line tension values are provided in the 
              units of $k_BT/\mathrm{nm}$. For each system, the results obtained analytically 
              within linearized PB theory and numerically within nonlinear as well as linear 
              PB theory are presented. As expected, the numerically obtained values for 
              $m=1$ in Eq.~\eqref{eq:7} (denoted as `Linear') agree with those obtained 
              analytically. As it is evident from the presented values, $\gamma_1$, $\gamma_2$, 
              and $\gamma_{1,2}$ are always negative for the systems considered here and 
              the linear theory always underestimates their absolute values. However, the 
              line tension $\tau$ appears to be the most sensitive to the variations in the 
              system parameters and to the type of theory used to calculate its value. It 
              can be positive as well as negative, decrease as well as increase and even 
              change sign while calculating within the linear and the nonlinear theories.}
\renewcommand{\arraystretch}{1.0} 
\newcolumntype{C}{>{\centering\arraybackslash}X}
\newcolumntype{R}{>{\raggedleft\arraybackslash}X}
\newcolumntype{L}{>{\raggedright\arraybackslash}X}
\centering
\begin{tabularx}{\textwidth}{LLCCCC}
\hline
\hline
& & $\gamma_1\,(k_BT/\mathrm{nm}^2)$ & $\gamma_2\,(k_BT/\mathrm{nm}^2)$ 
& $\gamma_{1,2}\,(k_BT/\mathrm{nm}^2)$ & $\tau\,(k_BT/\mathrm{nm})$ \\ 
\hline
\multirow{3}{*}{Water--lutidine} & Analytical & $-0.05116\phantom{0}$ & $-0.07780\phantom{0}$ & $-0.002621$ & $\phantom{-}0.03255\phantom{0000}$ \\
    \multirow{3}{*}{(Standard)} & Linear & $-0.05116\phantom{0}$ & $-0.07781\phantom{0}$ & $-0.002621$ & $\phantom{-}0.03255\phantom{0000}$ \\
                                & Nonlinear & $-0.06150\phantom{0}$ & $-0.1073\phantom{00}$ & $-0.002635$ & $\phantom{-}0.009874\phantom{000}$ \\ 
                          
 & & & & & \\
                          
\multirow{3}{*}{$\beta e\Psi_P=-5$} & Analytical & $-0.1421\phantom{00}$ & $-0.1751\phantom{00}$ & $-0.002621$ & $\phantom{-}0.03309\phantom{0000}$ \\
                                    & Linear & $-0.1421\phantom{00}$ & $-0.1751\phantom{00}$ & $-0.002621$ & $\phantom{-}0.03309\phantom{0000}$ \\
                                    & Nonlinear & $-0.2320\phantom{00}$ & $-0.3464\phantom{00}$ & $-0.002635$ & $-0.01676\phantom{0000}$ \\
                                                              
 & & & & & \\                                                              

\multirow{3}{*}{$\beta e\Psi_P=-1$} & Analytical & $-0.005684$ & $-0.01945\phantom{0}$ & $-0.002621$ & $\phantom{-}0.03200\phantom{0000}$ \\
                                    & Linear & $-0.005685$ & $-0.01945\phantom{0}$ & $-0.002621$ & $\phantom{-}0.03201\phantom{0000}$ \\
                                    & Nonlinear & $-0.005804$ & $-0.02113\phantom{0}$ & $-0.002635$ & $\phantom{-}0.02747\phantom{0000}$ \\

 & & & & & \\

\multirow{3}{*}{$\beta e\Psi_P=3$} & Analytical & $-0.05116\phantom{0}$ & $-0.01945\phantom{0}$ & $-0.002621$ & $\phantom{-}0.03091\phantom{0000}$ \\
                                   & Linear & $-0.05116\phantom{0}$ & $-0.01945\phantom{0}$ & $-0.002621$ & $\phantom{-}0.03091\phantom{0000}$ \\
                                   & Nonlinear & $-0.06150\phantom{0}$ & $-0.02113\phantom{0}$ & $-0.002635$ & $\phantom{-}0.01918\phantom{0000}$ \\

 & & & & & \\

\multirow{3}{*}{$\beta e\Psi_D=-1$} & Analytical & $-0.05116\phantom{0}$ & $-0.01945\phantom{0}$ & $-0.002621$ & $\phantom{-}0.03091\phantom{0000}$ \\
                                    & Linear & $-0.05116\phantom{0}$ & $-0.01945\phantom{0}$ & $-0.002621$ & $\phantom{-}0.03091\phantom{0000}$ \\
                                    & Nonlinear & $-0.06150\phantom{0}$ & $-0.02113\phantom{0}$ & $-0.002635$ & $\phantom{-}0.01918\phantom{0000}$ \\

 & & & & & \\

\multirow{3}{*}{$\beta e\Psi_D=0$} & Analytical & $-0.05116\phantom{0}$ & $-0.04377\phantom{0}$ & $\phantom{-}0\phantom{.000000}$ & $-0.000003198$ \\
                                   & Linear & $-0.05116\phantom{0}$ & $-0.04377\phantom{0}$ & $\phantom{-}0\phantom{.000000}$ & $-0.000003198$ \\
                                   & Nonlinear & $-0.06150\phantom{0}$ & $-0.05262\phantom{0}$ & $\phantom{-}0\phantom{.000000}$ & $-0.000003016$ \\

 & & & & & \\

\multirow{3}{*}{$\varepsilon=6.2/72$} & Analytical & $-0.05116\phantom{0}$ & $-0.02460\phantom{0}$ & $-0.001210$ & $-0.06384\phantom{0000}$ \\
                                                            & Linear & $-0.05116\phantom{0}$ & $-0.02463\phantom{0}$ & $-0.001211$ & $-0.06390\phantom{0000}$ \\
                                                            & Nonlinear & $-0.06150\phantom{0}$ & $-0.03415\phantom{0}$ & $-0.001224$ & $-0.07482\phantom{0000}$ \\

 & & & & & \\

\multirow{3}{*}{$I=0.085$} & Analytical & $-0.05116\phantom{0}$ & $-0.02460\phantom{0}$ & $-0.001210$ & $\phantom{-}0.06193\phantom{0000}$ \\
                                        & Linear & $-0.05118\phantom{0}$ & $-0.02461\phantom{0}$ & $-0.001210$ & $\phantom{-}0.06193\phantom{0000}$ \\
                                        & Nonlinear & $-0.06157\phantom{0}$ & $-0.03393\phantom{0}$ & $-0.001223$ & $\phantom{-}0.07061\phantom{0000}$ \\
                                        
 & & & & & \\

\multirow{3}{*}{Water--octanol} & Analytical & $-0.1670\phantom{00}$ & $-0.005242$ & $-0.001627$ & $\phantom{-}0.1478\phantom{00000}$ \\
                                & Linear & $-0.1677\phantom{00}$ & $-0.005220$ & $-0.001634$ & $\phantom{-}0.1455\phantom{00000}$ \\
                                & Nonlinear & $-0.2031\phantom{00}$ & $-0.01217\phantom{0}$ & $-0.002165$ & $\phantom{-}0.1321\phantom{00000}$ \\
\hline
\hline
\end{tabularx}
\label{Tab1}
\end{table*}

It is important to note that, the equilibrium 
separation between the particles is determined 
by the total interaction energy 
$\Delta\Omega(L)=\omega_{\gamma,1}(L)A_1+\omega_{\gamma,2}(L)A_2+\omega_{\tau}(L)\ell$; 
the curves in Fig.~\ref{Fig5} do not directly 
dictate it. However, they provide important 
informations regarding what one should expect 
for a given experimental set-up. For example, 
the surface interactions being always repulsive, 
inter-particle attraction can only come from 
the line part. As the plots suggest, the line 
part can either be monotonically attractive 
or it can change from repulsive to attractive 
beyond the position of a minimum at short 
separations. Even if it is monotonically 
attractive, at very short separations (below 
$L\approx 20\,\mathrm{nm}$) the surface parts 
usually dominate. Only in the case of small 
$\varepsilon$ values, the surface part is 
strongly suppressed. However, the line part 
itself becomes repulsive in this case (see 
Fig.~\ref{Fig5}(f)). The curves plotted 
in Figs.~\ref{Fig2}, \ref{Fig3} and \ref{Fig5} 
also suggest that if a minimum exists at a 
non-vanishing separation, it typically occurs 
between $L\approx 20\,\mathrm{nm}-40\,\mathrm{nm}$ 
(please note that $\kappa_1$ for Fig.~\ref{Fig3} 
is different than that in Figs.~\ref{Fig2} 
and \ref{Fig5}). Therefore, in view of these 
results, we conclude that overall electrostatic 
attraction can start to kick in at or above 
$L\approx 20\,\mathrm{nm}$ and the particles 
will arrange themselves with an inter-particle 
separation $L\gtrsim 20\,\mathrm{nm}$. It is 
worth mentioning that the relevance of the 
line part and the precise location of the 
minimum not only depend on the system parameters 
considered in Fig.~\ref{Fig5} but also on the 
surface areas exposed to the fluids and on 
the length of the three-phase contact line. 
However, from Fig.~\ref{Fig5}, one can easily 
say that the magnitude of the line part can 
be strongly enhanced by increasing the contrast 
in the ionic strengths in the two media (or 
equivalently, decreasing the ionic strength 
ratio $I=I_2/I_1$). On the other hand, for a 
given separation, the surface parts can be 
suppressed by decreasing the relative permittivity 
value, by reducing the surface potential, or 
by increasing the electrostatic screening. In 
fact, this is also evident if one compares 
the two experimental systems (water--lutidine 
and water--octanol) discussed previously.

What remain to be discussed are the $L$-independent 
interactions in Eqs.~\eqref{eq:3} or \eqref{eq:7}. 
Their values are listed in Table \ref{Tab1} for 
all the different system parameters considered 
here. As one can see, in all cases the numerically 
obtained values for $m=1$ in Eq.~\eqref{eq:7} 
(denoted as `Linear') agrees perfectly with the 
analytically obtained results (denoted as 
`Analytical') using Eqs.~\eqref{eq:9}, 
\eqref{eq:11}, and \eqref{eq:12}. As one can 
see from Table \ref{Tab1}, the surface tension 
$\gamma_1$ acting between the plates and 
medium ``1'' is always negative. However, its 
absolute value is higher within the nonlinear 
theory compared to the linear theory. Similar 
observations hold true for the surface tension 
$\gamma_2$ acting between the plates and 
medium ``2''. Both of them also increases with 
the absolute value of the difference between 
the surface potential $\beta e\Psi_P$ and the 
bulk potential $\beta e\Psi_{b,i}$. As one can 
infer from the fourth column, the absolute value 
of $\gamma_2$ decreases with decreasing 
$\beta e\Psi_D$, $\varepsilon$, and $I$ within 
both the nonlinear and the linear theories. As 
expected, the interfacial tension $\gamma_{1,2}$ 
(given in the fifth column) remains unaffected 
due to changes in the surface potential $\Psi_P$ 
and is insensitive to the sign of the Donnan 
potential $\beta e\Psi_D$. With decreasing 
$\varepsilon$ and $I$, the magnitude of 
$\gamma_{1,2}$ also decreases. For all the 
situations considered here, $\gamma_{1,2}$ is 
always negative except for $\beta e\Psi_D=0$ 
when it vanishes completely. The quantity 
that is most sensitive to changes in the 
system parameters is the line tension $\tau$ 
presented in the last column. It can be positive 
as well as negative. Not only that, it can 
decrease as well as increase and even change 
sign while calculating within the linear and 
the nonlinear theories. The values for the line 
tensions that we obtain are within the range 
$0.1\,\mathrm{pN}-1\,\mathrm{pN}$, which is 
in accordance with the values reported in the 
literature \cite{Mug02}.

We note that, although not shown here, the 
agreement of the results corresponding to $m=1$ 
in Eq.~\eqref{eq:7} with those obtained analytically 
within the linear theory have been checked for all 
the systems considered here. As another check, 
we also observe that with decreasing 
$\beta e\left|\Psi_P-\Psi_{b,i}\right|$ 
values, the difference between the results 
within the linear and the nonlinear theory diminishes.

\section{Conclusions}\label{IV}

To summarise, by using the classical DFT formalism, we have addressed 
the problem of the electrostatic interaction between a pair of identical 
particles with constant surface potentials sitting in close proximity of each 
other at a fluid-fluid interface. The particles could be either monometallic, 
or metal coated such as Janus particles with metallic caps facing each 
other or core-shell particles with metallic shells. Considering a simple yet 
reasonable model system, which exploits the short separation between 
the particles, we have solved the problem within both the linear and the 
nonlinear PB theory. Within the linear theory, closed-form analytical 
solutions are obtained for all the interaction parameters present in the 
system. For the nonlinear PB theory, the problem is solved numerically. 
While the results within the linear theory are mostly quantitatively inaccurate, 
they can be qualitatively reliable depending upon the system. The surface 
interaction energies between the two particles are always found to be 
repulsive and monotonically decaying. However, the line interaction 
energy varies non-monotonically with increasing separation between 
the particles and it can be attractive as well as repulsive. Not only that, 
contrary to common expectation, it is shown that the attractive 
line interaction can be strong enough to dominate over the sum of the 
surface parts and to dictate the nature of the overall electrostatic 
interaction for typical experimental systems. A comparison between two 
typical experimental setups illustrates that the importance of the line 
interaction enhances with increasing contrast in the bulk properties 
of the two fluids forming the interface. Therefore, special attention should 
be given to line interaction in the vast majority of experiments performed 
using liquids varying starkly in their bulk properties. We also provide results 
concerning the interactions such as surface, line, and interfacial tensions that 
do not depend on the separation distance between the particles. Finally, we 
note that the presented theoretical framework is based only on a simple 
minimization of the system free energy and can, in principle, be easily be 
applied to related interfacial problems dealing not necessarily with planar 
surfaces or interfaces. Therefore, on one hand, we expect our study to directly contribute 
towards improving the understanding and modeling of interfacial monolayers of 
colloidal particles with constant surface potentials. On the other hand, we expect 
it to serve as a useful reference for future theoretical investigations of related 
problems.

\begin{acknowledgments}
Helpful discussions with Markus Bier are gratefully acknowledged.
\end{acknowledgments}
\smallskip
This article may be downloaded for personal use only. Any other use requires 
prior permission of the authors and AIP Publishing. This article appeared in 
the Journal of Chemical Physics and may be found at 
\href{https://aip.scitation.org/doi/10.1063/5.0013298}{J.\ Chem.\ Phys.\ \textbf{153}, 044903 (2020)}.


\end{document}